\documentstyle[12pt]{article}
\newcommand{\nc}{\newcommand}
\nc{\renc}{\renewcommand}
\def\lsim{\; \raise0.3ex\hbox{$<$\kern-0.75em
      \raise-1.1ex\hbox{$\sim$}}\; }
\def\gsim{\; \raise0.3ex\hbox{$>$\kern-0.75em
      \raise-1.1ex\hbox{$\sim$}}\; }
\def\lae{\;^{<}_{\sim} \;}  

\def\GeV{{\rm\ GeV}}

\nc{\advp}[3]{{\it  Adv.\ in\ Phys.\ }{{\bf #1} {(#2)} {#3}}}
\nc{\annp}[3]{{\it  Ann.\ Phys.\ (N.Y.)\ }{{\bf #1} {(#2)} {#3}}}
\nc{\apl}[3]{{\it  Appl. Phys. Lett. }{{\bf #1} {(#2)} {#3}}}
\nc{\apj}[3]{{\it  Ap.\ J.\ }{{\bf #1} {(#2)} {#3}}}
\nc{\apjl}[3]{{\it  Ap.\ J.\ Lett.\ }{{\bf #1} {(#2)} {#3}}}
\nc{\app}[3]{{\it Astropart.\ Phys.\ }{{\bf #1} {(#2)} {#3}}}
\nc{\cmp}[3]{{\it  Comm.\ Math.\ Phys.\ }{{ \bf #1} {(#2)} {#3}}}
\nc{\cqg}[3]{{\it  Class.\ Quant.\ Grav.\ }{{\bf #1} {(#2)} {#3}}}
\nc{\epl}[3]{{\it  Europhys.\ Lett.\ }{{\bf #1} {(#2)} {#3}}}
\nc{\ijmp}[3]{{\it Int.\ J.\ Mod.\ Phys.\ }{{\bf #1} {(#2)} {#3}}}
\nc{\ijtp}[3]{{\it Int.\ J.\ Theor.\ Phys.\ }{{\bf #1} {(#2)} {#3}}}
\nc{\jmp}[3]{{\it  J.\ Math.\ Phys.\ }{{ \bf #1} {(#2)} {#3}}}
\nc{\jpa}[3]{{\it  J.\ Phys.\ A\ }{{\bf #1} {(#2)} {#3}}}
\nc{\jpc}[3]{{\it  J.\ Phys.\ C\ }{{\bf #1} {(#2)} {#3}}}
\nc{\jap}[3]{{\it J.\ Appl.\ Phys.\ }{{\bf #1} {(#2)} {#3}}}
\nc{\jpsj}[3]{{\it J.\ Phys.\ Soc.\ Japan\ }{{\bf #1} {(#2)} {#3}}}
\nc{\lmp}[3]{{\it Lett.\ Math.\ Phys.\ }{{\bf #1} {(#2)} {#3}}}
\nc{\mpl}[3]{{\it  Mod.\ Phys.\ Lett.\ }{{\bf #1} {(#2)} {#3}}}
\nc{\np}[3]{{\it  Nucl.\ Phys.\ }{{\bf #1} {(#2)} {#3}}}
\nc{\npps}[3]{{\it  Nucl.\ Phys.\ Proc.\ Suppl.\ }{{\bf #1} {(#2)} {#3}}}
\nc{\pr}[3]{{\it Phys.\ Rev.\ }{{\bf #1} {(#2)} {#3}}}
\nc{\pra}[3]{{\it  Phys.\ Rev.\ A\ }{{\bf #1} {(#2)} {#3}}}
\nc{\prb}[3]{{\it  Phys.\ Rev.\ B\ }{{{\bf #1} {(#2)} {#3}}}}
\nc{\prc}[3]{{\it  Phys.\ Rev.\ C\ }{{\bf #1} {(#2)} {#3}}}
\nc{\prd}[3]{{\it  Phys.\ Rev.\ D\ }{{\bf #1} {(#2)} {#3}}}
\nc{\prl}[3]{{\it Phys.\ Rev.\ Lett.\ }{{\bf #1} {(#2)} {#3}}}
\nc{\pl}[3]{{\it  Phys.\ Lett.\ }{{\bf #1} {(#2)} {#3}}}
\nc{\prep}[3]{{\it Phys.\ Rep.\ }{{\bf #1} {(#2)} {#3}}}
\nc{\prsl}[3]{{\it Proc.\ R.\ Soc.\ London\ }{{\bf #1} {(#2)} {#3}}}
\nc{\ptp}[3]{{\it  Prog.\ Theor.\ Phys.\ }{{\bf #1} {(#2)} {#3}}}
\nc{\ptps}[3]{{\it  Prog\ Theor.\ Phys.\ suppl.\ }{{\bf #1} {(#2)} {#3}}}
\nc{\physa}[3]{{\it  Physica\ A\ }{{\bf #1} {(#2)} {#3}}}
\nc{\physb}[3]{{\it  Physica\ B\ }{{\bf #1} {(#2)} {#3}}}
\nc{\phys}[3]{{\it Physica\ }{{\bf #1} {(#2)} {#3}}}
\nc{\rmp}[3]{{\it  Rev.\ Mod.\ Phys.\ }{{\bf #1} {(#2)} {#3}}}
\nc{\rpp}[3]{{\it Rep.\ Prog.\ Phys.\ }{{\bf #1} {(#2)} {#3}}}
\nc{\sjnp}[3]{{\it Sov.\ J.\ Nucl.\ Phys.\ }{{\bf #1} {(#2)} {#3}}}
\nc{\spjetp}[3]{{\it Sov.\ Phys.\ JETP\ }{{\bf #1} {(#2)} {#3}}}
\nc{\yf}[3]{{\it Yad.\ Fiz.\ }{{\bf #1} {(#2)} {#3}}}
\nc{\zetp}[3]{{\it Zh.\ Eksp.\ Teor.\ Fiz.\  }{{\bf #1}  {(#2)} {#3}}}
\nc{\zp}[3]{{\it Z.\ Phys.\ }{{\bf #1} {(#2)} {#3}}}
\nc{\ibid}[3]{{\sl ibid.\ }{{\bf #1} {#2} {#3}}}
\newlength{\undereqskip}
\nc{\be}[1]{\begin{equation} \mbox{$\label{#1}$}}
\nc{\bea}[1]{\begin{eqnarray} \mbox{$\label{#1}$}}
\nc{\eea}{\vspace{\undereqskip}\end{eqnarray}}
\nc{\ee}{\vspace{\undereqskip}\end{equation}}
\nc{\nn}{\nonumber \\*}
\nc{\eqs}[2]{\mbox{Eqs.~(\ref{#1},\,\ref{#2})}}
\nc{\eq}[1]{\mbox{Eq.~(\ref{#1})}}
\nc{\etal}{\mbox{\it et al. }}

\begin{document}
{\title{\vskip-2truecm{\hfill {{\small HIP-1998-63/TH\\
	}}\vskip 1truecm}
{\bf Dark Matter from Unstable B-balls$^1$}}
{\author{
{\sc  Kari Enqvist$^{2}$}\\
{\sl\small Department of Physics and Helsinki Institute of Physics}
\\
{\sl\small P.O. Box 9,
FIN-00014 University of Helsinki,
Finland}}
}
\maketitle
\vspace{1cm}
\begin{abstract}\noindent
The spectrum of MSSM admits solitons carrying baryonic charge, or B-balls.
In an inflationary universe they can be produced in significant
numbers by a break-up of a scalar condensate along the flat directions.
It is shown that
if SUSY breaking is mediated to the observable sector by gravity,
B-balls are unstable but decay to baryons and LSPs
typically well below the electroweak
phase transition. It is argued that B-balls could be the source of
most baryons and cold dark matter in the universe, with their
number densities related by $n_{LSP}\simeq 3n_{B}$. For B-balls to
survive thermalization, the reheating temperature after inflation
should be less than about $10^{3}$ GeV.
\end{abstract}
\vfil
\footnoterule
{\small $^1$ Invited talk
at DARK98 conference, Heidelberg, Germany, July 20-25
\vskip-2pt\noindent
$^2$enqvist@pcu.helsinki.fi};
\thispagestyle{empty}
\newpage
\setcounter{page}{1}
\section{Introduction}
A Q-ball is a stable, charge Q non-topological 
soliton in a scalar field theory 
with a spontaneously broken global $U(1)$ symmetry \cite{cole}.
The  Q-ball solution arises provided the scalar potential $V(\phi)$
is such that $V(\phi)/\vert\phi\vert^2$ has a minimum at non-zero $\phi$.
Although not found in the Standard Model, 
the  spectrum of the MSSM has  Q-balls
 which carry baryonic charge and are therefore called 
B-balls \cite{cole2,ks1}. 
In a cosmological scenario which includes
inflation they can be copiously produced by the breakdown of scalar 
condensates along the flat directions of the MSSM \cite{ks2,bbb1}.
The properties  of the MSSM Q-balls will depend  upon 
the scalar potential
 associated with the condensate scalar, which in turn depends upon 
the SUSY breaking mechanism and on the order d
at which the non-renormalizable
terms lift the degeneracy of the potential; examples are the 
$H_{u}L$-direction with d=4 and $u^{c}d^{c}d^{c}$-direction
with d=6 \cite{drt}. If
SUSY breaking occurs at low energy scales, via gauge
 mediated SUSY breaking \cite{gmsb}, Q-balls will be stable \cite{ks2,dks}. 
This is so because for large enough $\phi$, the scalar potential is
essentially flat and the resulting Q-ball will be very tightly bound
with energy that grows as $\sim Q^{1/4}$.
Stable B-balls 
could have a wide range of astrophysical, 
experimental and practical 
implications, extensively discussed 
in references [3,4,8-12]; for instance, stable B-balls could account for cold 
dark matter \cite{ks2}. 

In the case of gravity-mediated
breaking, studied in \cite{bbb1,bbb2}, 
B-balls are unstable. However, if they can survive thermalization,
they are typically
long-lived enough to decay much after the electroweak phase transition, 
leading to a variant of the Affleck-Dine (AD) mechanism 
\cite{ad} known as B-ball Baryogenesis (BBB).
 
The requirement that B-balls can survive thermalization implies that 
B-balls in R-parity conserving models originate from 
a d=6 AD condensate and imposes
 an upper bound on the reheating temperature of $10^{3-5}\GeV$ 
\cite{bbb1,bbb2}. 
Such B-balls can protect a B asymmetry originating in the AD 
condensate from the effects 
of additional B-L violating interactions, 
which would otherwise wash out the B asymmetry when combined with 
anomalous B+L violation \cite{bbb1}. In addition, if the reheating 
temperature is sufficiently low so that the B-balls decay below the 
freeze-out temperature of the lightest SUSY particle (LSP
), then cold dark matter can mostly come from B-ball decays
rather than from thermal relics. This
opens up the possibility of relating the number density of dark 
matter particles to that of baryons, allowing for an explanation 
of their observed similarity for the case of dark matter particles 
with weak scale masses \cite{bbb2, bbbdm}.

\section{Unstable B balls}

If SUSY 
breaking occurs via the supergravity hidden sector, the potential
is not flat, but nevertheless radiative corrections to the $\phi^{2}$-type
 condensate potential allow B-balls to form \cite{bbb1,bbb2}.
Along the d=6 $u^cu^cd^c$ flat direction it reads
\be{pot}
V_6\simeq m^{2}_S |\phi|^{2} 
+ \frac{\lambda^{2}|\phi|^{10}
}{M_{P}^{6}} + \left( \frac{A_{\lambda} 
\lambda \phi^{6}}{M_{P}^{3}} + h.c.\right)    
~,\ee
where $\lambda$ and $A$ are coupling constants and 
the SUSY breaking mass $m_S^2\simeq m_0^2[1+K\log(\vert\phi\vert^2/\phi_0^2)]$,
where $\phi_0$ is the reference point and $K$ a negative constant (and
arises mainly because of  gaugino loops), 
decreases as $\phi$ grows, thus satisfying the requirement that
$V(\phi)/\vert\phi\vert^2$ has a minimum at non-zero $\phi$.
The potential is stabilized by the non-renormalizable term
so that inside the condensate the squark 
field takes the value $\langle\phi\rangle\simeq
4\times 10^{14}$ GeV. The decreasing of the effective mass term is
also responsible for the growth of any initial perturbation. In particular,
there are perturbations in the condensate field inherited from the
inflationary period. As was discussed in ref. 13, 
these will grow and become non-linear when
$H=H\simeq 2\vert K\vert m_S\alpha^{-1}$,
where $\alpha\simeq -log(\delta\phi_0(\lambda_0)/\phi_0)$
with $\lambda_{0}$  the length scale of the perturbation at $H \simeq m_S$,
and $\phi_{0}$ is the value of $\phi$ when the condensate oscillations begin.
The charge of the condensate lump is determined by the baryon asymmetry 
of the Universe at the time $H=H_i$ 
and the initial size of the perturbation when it goes non-linear 
\cite{bbb1,bbb2}.
The baryon asymmetry of the Universe at a given value of $H$ during 
inflaton oscillation
domination is given by 
\be {ba}
n_{B} = \left( \frac{\eta_{B}}{2 \pi} \right) 
\frac{H^{2} M_{Pl}^{2}}{T_{R}}  \simeq 
1.6 \times 10^{18} H^{2} \left(\frac{10^{9}}{T_{R}}\right)     
~,\ee
where we have taken the baryon to entropy ratio to be $\eta_{B} \simeq 10^{-10}$. 
It can be shown \cite{bbb2} 
that the charge in the initial condensate lump is given by 
\be{charge} 
B = \frac{4 \pi^{3}}{3 \sqrt{2}} 
\frac{\eta_{B} |K|^{1/2} M_{pl}^{2}}{m_S \alpha^{2} T_{R}} 
= 2 \times10^{15} |K|^{1/2} 
\left(\frac{100 \GeV}{m_S}\right)
\left(\frac{10^{9} \GeV}{T_{R}}\right)
\left(\frac{40}{\alpha}\right)^{2}
~\ee
where we have used $\alpha(\lambda_{0}) = 40$ as a typical value.

Once the d=6 AD condensate collapses, a fraction $f_{B}$ of the total 
B asymmetry ends up in the form of B-balls.
The formation of B-balls from the AD condensate can be shown to be 
generally effective \cite{bbb2} if the charge density inside the initial lump is
small enough; this can be translated to a condition on the reheating
temperature which reads
\be{tr} 
T_{R} \gsim \frac{\eta_{B} m M_{Pl}^{2}}{8 \pi \phi_{0}^{2}}
= 0.23 \left(\frac{m_S}{100 \GeV}\right)
 ~.\ee

\begin{figure}
\leavevmode
\centering
\vspace*{72mm} 
\includegraphics{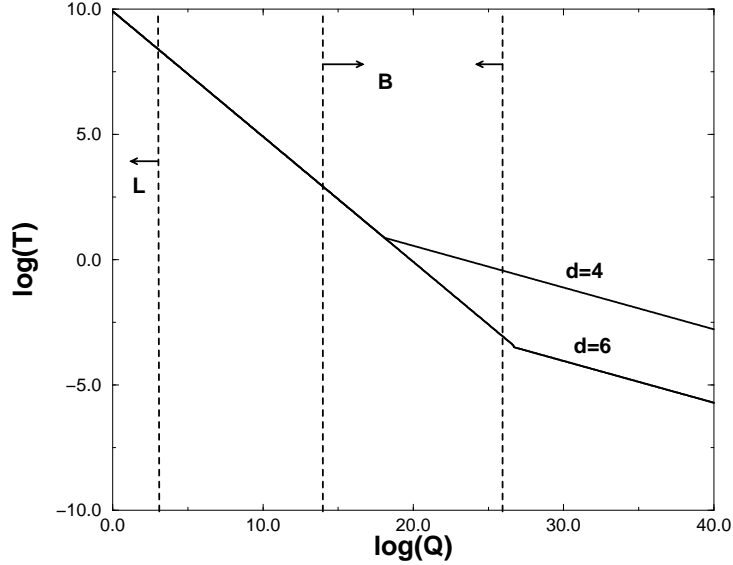}   
\caption{Q-ball decay temperature $T$ vs. the charge $Q$ for d=4 and d=6
Q-balls. The regions
where L-balls and B-balls exist are also indicated.}
\label{kuva2}       
\end{figure} 

After the formation B-balls could be dissociated by the bombardment of
thermal particles, or dissolve by charge escaping from the outer layers.
Both problems can be avoided \cite{bbb2} 
provided $T_{R} \lsim 10^3-10^5$ GeV for 
$\vert K\vert$ in the range 0.01 to 0.1. It then follows that
the surviving B-balls must have very large charges, $B\gsim 10^{14}$. 
The decay rate of the
B-ball also depends on its charge and takes place  at a temperature
\be{decayT} 
T_{d} \simeq 0.01 
\left(\frac{f_{s}}{f_{B}}\right)^{1/2}
\left(\frac{0.01}{|K|}\right)^{3/4}
\left(\frac{m}{100 \GeV}\right)
\left(\frac{T_R}{1 \GeV}\right)^{1/2}
\GeV
~,\ee
where $m$ is the B-ball squark mass and $f_{s}$ is the possible enhancement factor if the squarks can decay to a pair of scalars rather
than to final states with two fermions; we have estimated $f_{s} \simeq 10^{3}$ \cite{bbb2}. 
($f_{B}$ and $T_{d}$ are the only B-ball parameters which enter into the determination of the LSP density from B-ball decay). 
For example, with $T_{R} \simeq 1 \GeV$, as suggested by the 
d=6 AD mechanism, 
and with $f_{B}$ in the range 0.1 to 1 
(in accordance with an argument \cite{bbb2} 
that B-ball formation from an AD condensate is likely to be 
very efficient, although the numerical value of $f_{B}$ is not yet known), 
$T_{d}$ will generally be in the range 1 MeV to 1 GeV. 

Assuming $f_s=1$, the 
decay temperature is depicted in Fig. 1 as a function of the Q-ball charge 
for both thin and thick-wall Q-balls, which have different surface areas
(d=6 B-ball is of the thick-wall variety \cite{bbb1}).  As can be seen, for 
$B\gsim 10^{14}$, B-balls will indeed decay well below the electroweak
phase transition temperature, providing a new source of baryon
asymmetry not washed away by sphaleron interactions.
The only requirement is relatively low reheating temperature after inflation,
typically of the order of 1 GeV, which is in fact also
implied by the 
observed baryon asymmetry when the CP violating phase responsible for the 
baryon asymmetry is of the order of 1 \cite{bbb2}. 
A low reheating temperature can be achieved in the currently popular
D-term inflation models \cite{kmr,bbbd} as a consequence of R-symmetries
needed to protect the flatness of the inflaton potential.


\section{Neutralinos from B balls}
When the B-ball decays, for each unit of baryon number about 3 units
of R-parity will also be produced (B-ball being essentially a condensate
made of squarks). As discussed in the previous section, this will
typically happen at or below
the LSP freeze-out temperature 
$T_f\simeq m_{LSP}/20$ \cite{susydm,jdm1} .
The neutralino density will then consist of a possible thermal relic component, $n_{relic}(T)$, and a 
component from B-ball decays, $n_{BB}(T)$. The value of 
$n_{BB}(T)$ will depend upon whether or not the 
LSPs from B-ball decay can subsequently annihilate. 
The upper limit on $n_{LSP}(T)$ from 
annihilations is given by
\be{ann} n_{LSP}(T) \lae n_{limit}(T) \equiv 
\left( \frac{H}{<\sigma v>_{ann}} \right)_{T}  ~,\ee
where $<\sigma v>_{ann}$ is the thermal average of the 
annihilation cross-section times the 
relative velocity of the LSPs, which can be generally written in the form
$<\sigma v>_{ann} = a + b T / m_{{\rm LSP}}$ \cite{susydm}.
If $n_{LSP}(T) \lae n_{limit}(T)$, and if 
the B-ball formation efficiency $f_{B}$ is not 
too small compared with 1, 
there will be a natural similarity between the number density of 
LSPs and that of the baryons. 
Otherwise the annihilation of neutralinos will suppress the 
number density of LSPs relative to that 
of the baryons, although we will still have an interesting 
non-thermal neutralino relic density.

If the reheating temperature is much less than $T_{f}$, there will be essentially no
thermal relic background of LSPs, since the additional entropy released during the inflaton matter domination period will strongly suppress the thermal relic density by a factor $(T_{R}/T_{f})^{5}$. The present direct experimental bound on the LSP mass, valid for any
$\tan\beta$ (but assuming $m_{\tilde\nu}\ge 200$ GeV),
is $m_{{\rm LSP}}\ge 25$ GeV \cite{lspbound}.
If one assumes the MSSM with universal soft SUSY breaking masses and unification, LEP results
can be combined to yield an excluded region in the 
$(m_{{\rm LSP}}, m_{\tilde l_R})$-plane \cite{lep1}. In the case of $\tilde e_R$, which
provides the most stringent bound, the excluded region is roughly
parametrized by $m_{{\rm LSP}}\lsim 0.95 m_{\tilde e_R}$ for $45\GeV\lsim
m_{\tilde e_R}
\lsim 78$ GeV (this result holds for ${\rm tan}\beta=2$ and $\mu=-200$
GeV) \cite{lep1}. Therefore the LSP freeze-out 
temperature is expected to be greater than about 1-2 GeV.
Thus there are two possibilities, depending on 
$T_R$ and $T_f$: either the
LSP cold dark matter density, 
$\Omega_{\rm LSP}$, will be given
solely by the  LSP density which originated from the B-ball decay, which
we denote by $\Omega_{\rm BB}$,
or there will also be a relic density so that $\Omega_{\rm LSP}=\Omega_{\rm BB}
+\Omega_{\rm relic}$.

             Assuming that $n_{LSP}(T_{d}) \lae n_{limit}(T_{d})$, the LSP density from B-ball
 decays will be given simply by
 \be{nd} 
n_{BB} = 3 f_{B} n_{B}     ~.\ee
Thus the B-ball produced LSP and baryonic densities will be related by
\be{bdm} \frac{\Omega_{B}}{\Omega_{BB}} 
= {m_N \over 3 f_{B}m_{\rm LSP}} ~.
\ee
B generation via the AD mechanism requires inflation \cite{ad,kmr}, and although varieties of inflationary
models exist with $\Omega_{\rm tot}<1$, let us
nevertheless adopt the point of view that
inflation implies $\Omega_{\rm tot}=1$ to a high precision. One may then write
\bea{omega1}
\Omega_{\rm tot}&=&\Omega_0+\Omega_{\rm LSP}+\Omega_B\nn
&=&\Omega_0+\Omega_{\rm relic}+\left({3 f_{B}m_{\rm LSP}\over m_N}
+1\right) \Omega_B =1~,
\eea
where $\Omega_0$ includes the hot dark matter (HDM) 
component and/or a possible cosmological constant.
Therefore $\Omega_{B}$ is fixed by $\Omega_0$,  $f_{B}$ 
and $m_{{\rm LSP}}$ together with the MSSM parameters entering into the annihilation rate. Applying nucleosynthesis bounds \cite{sarkar} on
$\Omega_{B}$ then gives constraints on these parameters. Note that, so long as LSP annihilations 
after B-ball decay can be neglected, the resulting LSP density is independent of $T_{d}$.

\begin{figure}[t]
\leavevmode
\centering
\vspace*{95mm} 
\includegraphics{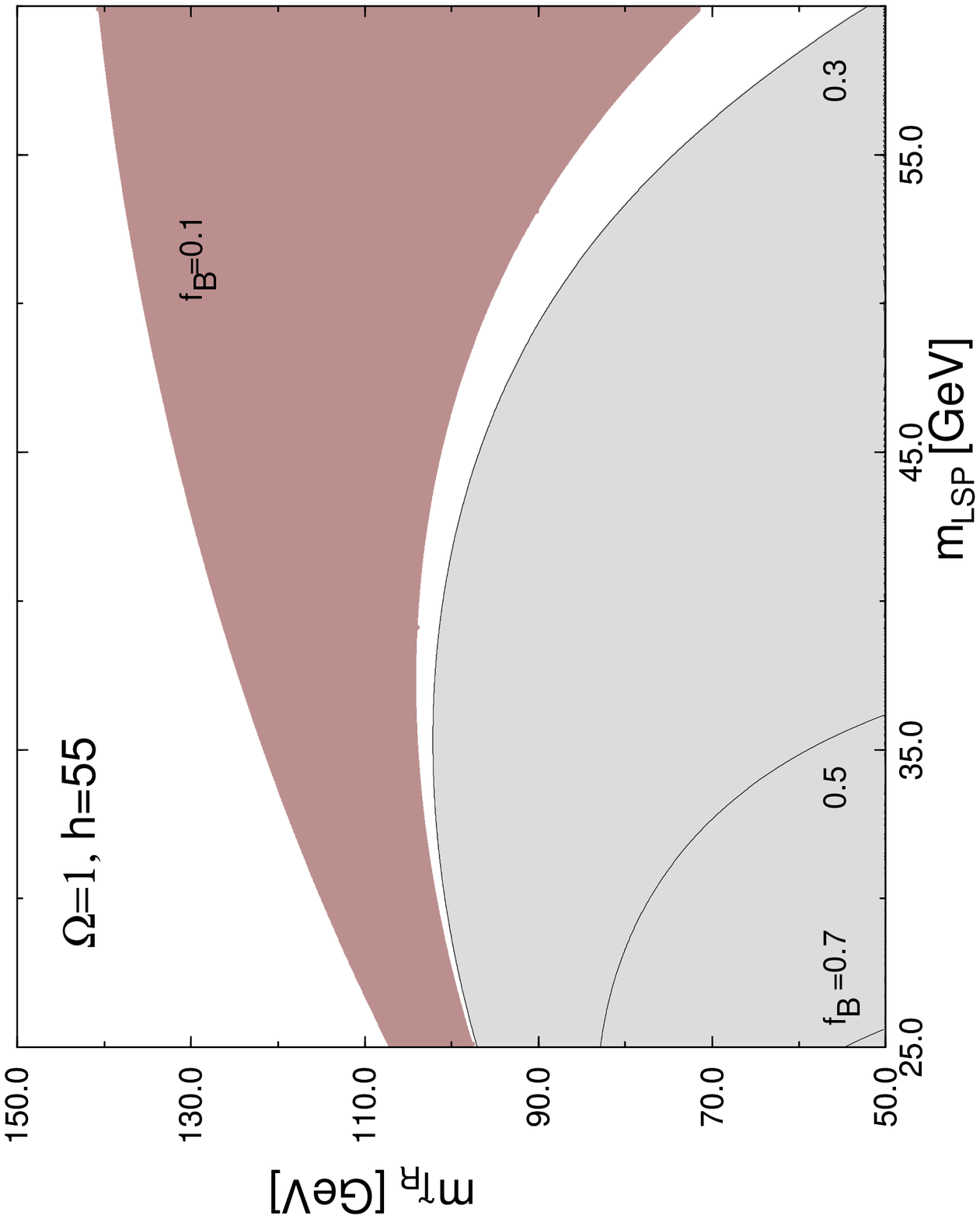}
\includegraphics{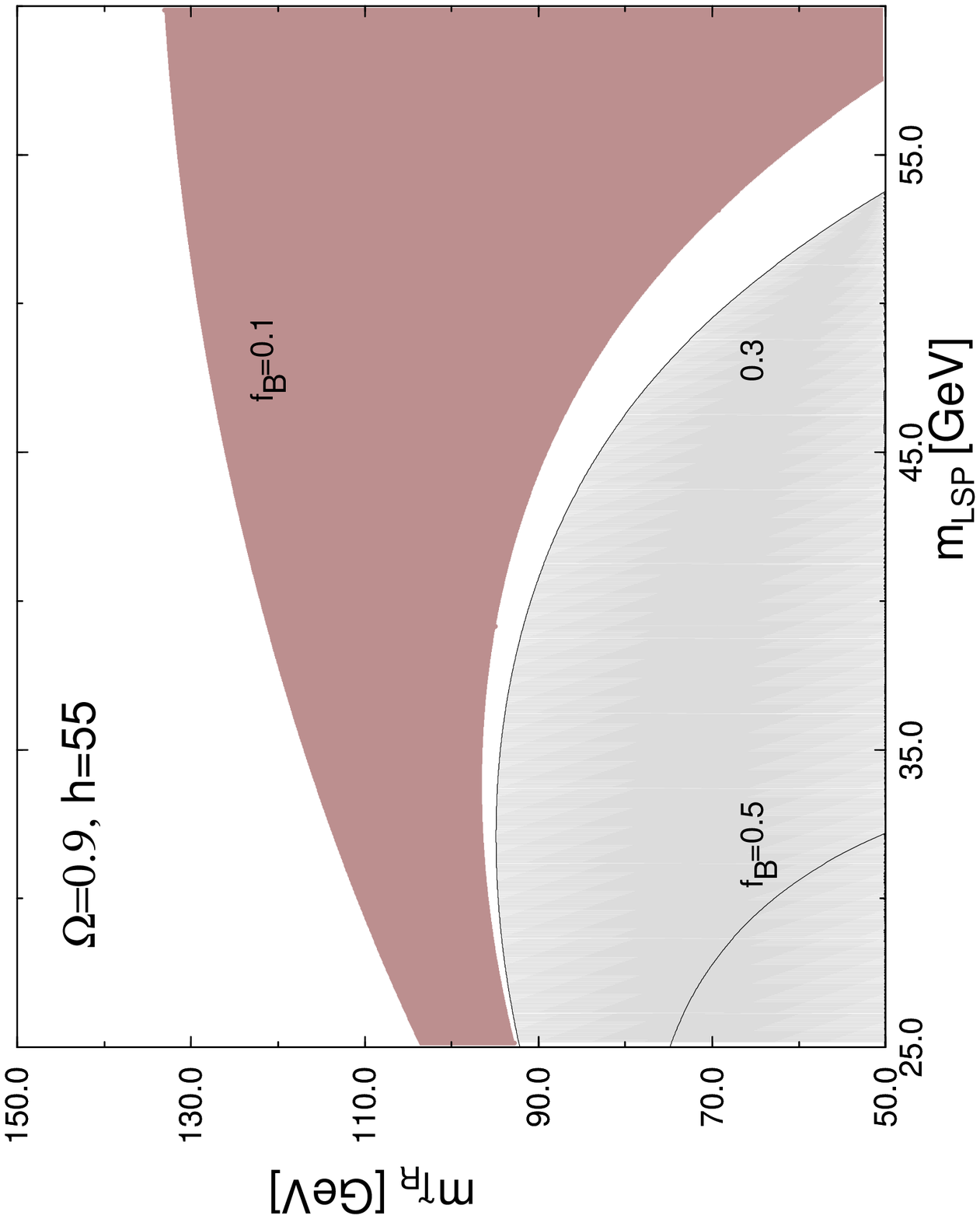}   
\includegraphics{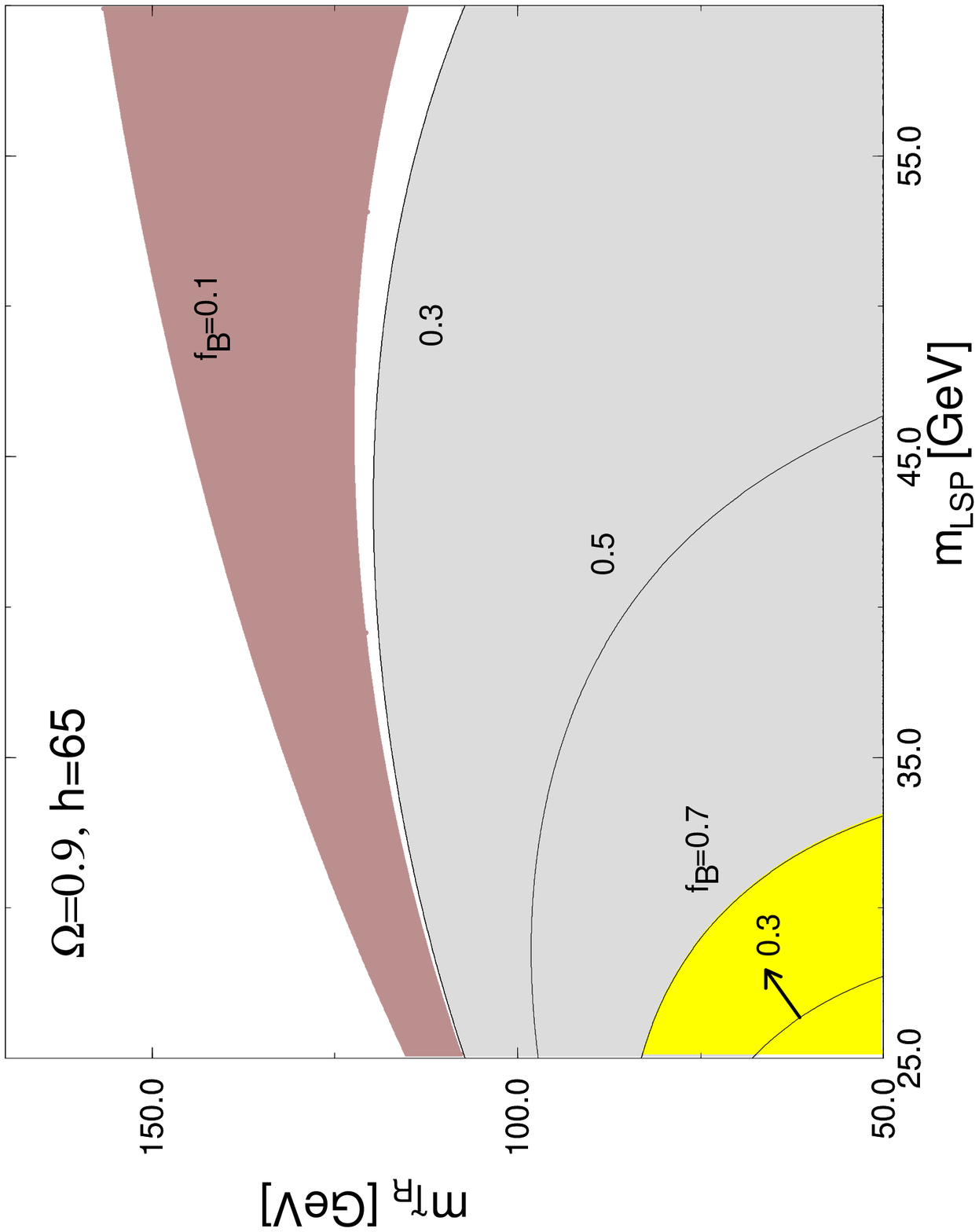}
\includegraphics{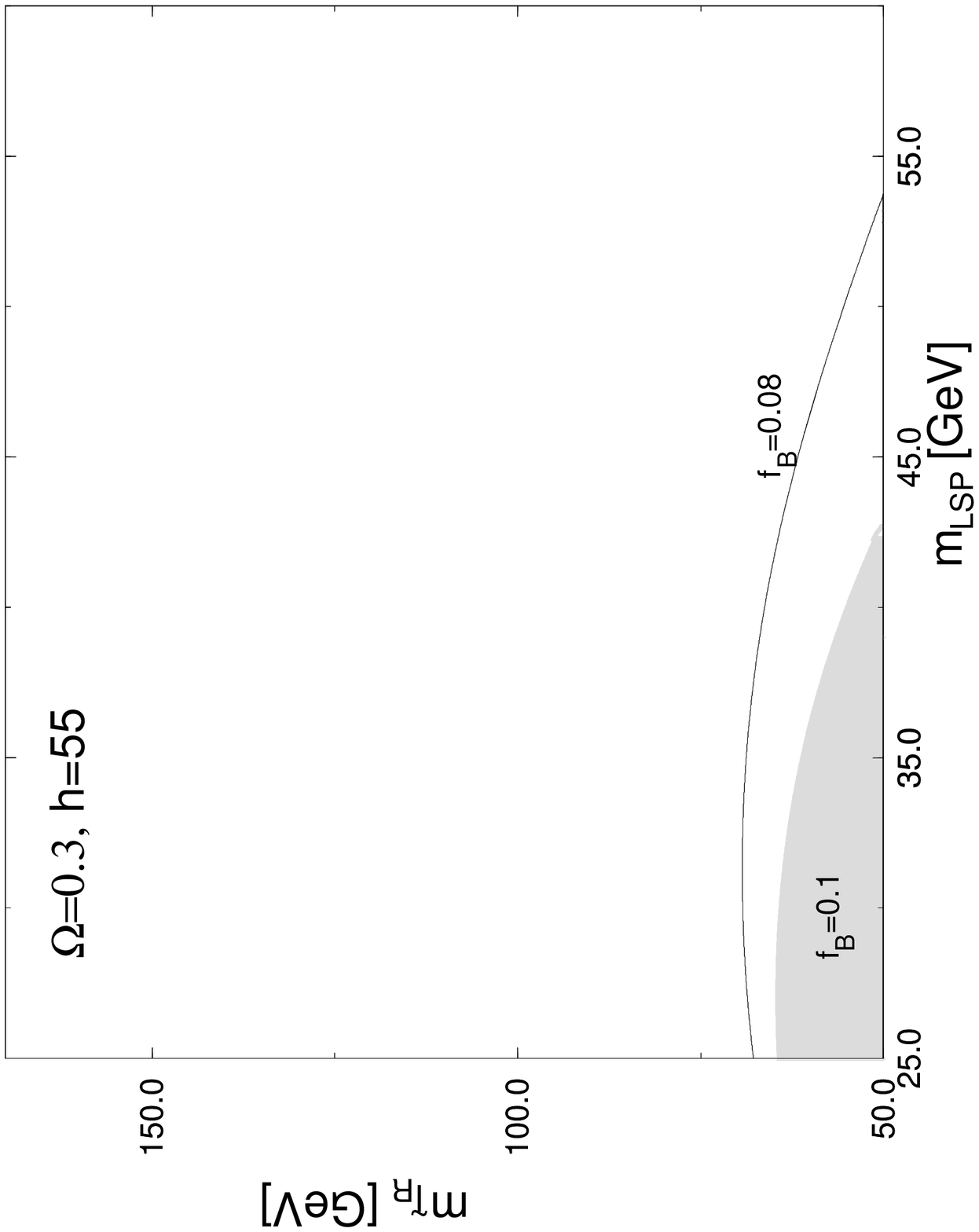}
\caption{The allowed regions in the $(m_{\rm LSP},m_{\tilde l_R})$-plane
for different values of the total CDM density $\Omega$ and 
the Hubble parameter $h$.}
\label{kuva2}       
\end{figure}

Let us first consider the case where the thermal relic density $\Omega_{relic}$
is negligible. This would be true if $T_R$ was sufficiently small compared with the freeze-out temperature $T_f$.  We then obtain 
the limit \cite{bbbdm}
\be{b1} 76.9 (1-\Omega_0) h^{2} - 1 \lsim \frac{3 m_{{\rm LSP}}f_{B}}{m_{N}} 
 \lsim  208.3 (1-\Omega_0) h^{2} - 1~.\ee
With $\Omega_0 = 0$ this would result in a bound on the LSP mass given by
\be{b2} 
3.8 
f_{B}^{-1} \GeV 
\lsim m_{{\rm LSP}}
\lsim 
29 
 f_{B}^{-1} \GeV  ~,
\ee
where we have used $0.4 \lae h \lae 0.65$.
If $f_{B} = 1$ this would be only marginally compatible with present experimental constraints and then only if we do not consider universal soft SUSY breaking masses.
Larger values of $\Omega_0$ impose even tighter bounds on $m_{{\rm LSP}}$, requiring $f_{B} 
< 1$. Therefore, in the absence of annihilations after 
B-ball decays, LSP dark matter from B-balls is likely to be compatible with nucleosynthesis bounds only if a significant fraction of the baryon asymmetry exists outside the B-balls. 
Reasonable values of $f_{B}$ can, however, accomodate an interesting range of LSP masses; for 
example, values in the range 0.1 to 1  
allow LSP masses as large as 290 GeV. $f_{B}$ can be calculated theoretically, but this requires an analysis of the non-linear evolution of the unstable AD condensate.
The comparison of the theoretical value with the dark matter constraints 
will be an important test of this scenario.

       Let us next consider the case with $T_R>T_f$. In this case 
there will be a significant thermal relic density 
and we can use nucleosynthesis 
bounds on $\Omega_{B}$ to constrain the masses of the particles 
responsible for the LSP annihilation cross-section. 
The constraints will depend on the identity of the LSP and the masses of the particles 
entering the LSP annihilation cross-section. In general, this would require
 a numerical analysis of the renormalization group equations for the SUSY particle spectrum.
However, for the case of universal scalar and gaugino masses at a large scale, the LSP 
is likely to be mostly bino and the lightest scalars are likely to be the right-handed sleptons \cite{bino}. 
This is consistent with the requirement that the LSP does not have a large coupling to the Z boson, which would otherwise efficiently annihilate 
away the thermal relics. However, there will be a small, model-dependent Higgsino component which will be important for LSP masses close to the Z pole. For LSP masses away from this pole, it will be a reasonable approximation
to treat the LSP as a $pure$ bino,  
although the possible suppression of the thermal relic 
density around the Z pole 
due to a Higgsino component and the subsequent weakening of 
MSSM constraints should be kept in mind. 

        For the case of a pure bino, the largest contribution to the annihilation cross-section 
comes from
$t$-channel $\tilde l_R$ exchange in ${\chi}{\rm \chi}\to l^+l^-$ \cite{bino}. 
In that case one finds \cite{bino}
\be{binos}
\Omega_{\rm relic}h^2={\Sigma^2\over M^2m_{\rm LSP}^2}
\left[\left(1-{m_{\rm LSP}^2\over \Sigma}\right)^2
+{m_{\rm LSP}^4 \over \Sigma^2}\right]^{-1}~,
\ee
where $M\simeq 1$ TeV and $\Sigma=m_{\rm LSP}^2+m_{\tilde l_R}^2$. 
Plugging this into \eq{omega1}
and using the range of $\Omega_B$ allowed by nucleosynthesis \cite{sarkar}, 
one may obtain \cite{bbbdm} allowed ranges
in the $(m_{\rm LSP},m_{\tilde l_R})$-plane. 
These are demonstrated in Fig. 2 for different values of
$\Omega_0$ and $h$. 

\begin{figure}[t]
\leavevmode
\centering
\vspace*{85mm} 
\includegraphics{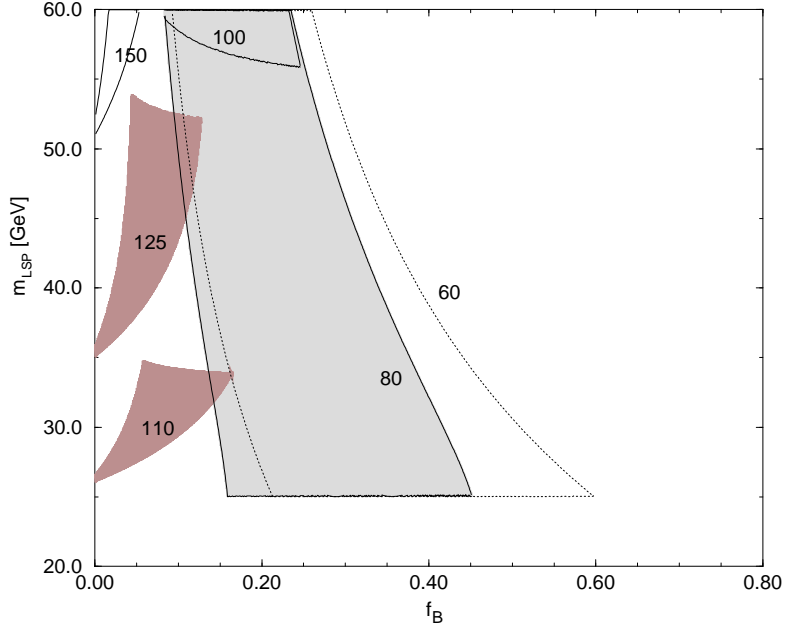}
\caption{The allowed region in the $(f_B, m_{\rm LSP})$-plane
for fixed $m_{\tilde l_R}$, assuming that
the total CDM density $\Omega=0.9$ and 
the Hubble parameter $h=0.65$.}
\label{kuva3}       
\end{figure} 

In the conventional MSSM case \eq{binos} would imply that both $m_{\rm LSP}$
and $m_{\tilde l_R}$ should be less than about 200 GeV. 
Because of the added B-ball contribution
a more stringent constraint follows in the present case. 
If the reheating temperature is
larger than the LSP freeze-out temperature, and if we consider the range $0.1 \lae f_B \lae 1$ to be the most likely, we may conclude that 
only a {\it very light
sparticle spectrum} is consistent with $\Omega = 1$; this is so in particular 
if there is a
cosmological constant with $\Omega_0\simeq 0.7$, as suggested by recent 
supernova 
studies \cite{SN}. In any case, it is evident that in the case $T_R>T_f$ 
one obtains significant constraints on the B-ball formation efficiency from MSSM constraints. 
This is demonstrated in Fig. 3 for the case of 
$\Omega_0=0.1$,
where the allowed regions in the $(f_B, m_{\rm LSP})$-plane for 
fixed values of $m_{\tilde l_R}$ are plotted. 
As can be seen, in the case of $T_R>T_f$ dark matter constrains $f_B$ to be 
less than about 0.6. If the SUGRA-based LEP limit 
 $m_{{\rm LSP}}\lsim 0.95 m_{\tilde e_R}$ ($45\GeV\lsim 
m_{\tilde e_R}\lsim 78$ GeV)
 is implemented \cite{lep1}, the limit on $f_B$
would be even lower. This serves to emphasize the need for an accurate
 theoretical determination of  
$f_B$.

\section{Conclusions}

 It is possible to reach only very broad conclusions 
about the sparticle spectrum at present, as the 
B-ball decay parameters $f_{B}$ and $T_{d}$ are unknown. However,
 both $f_{B}$ and $T_{d}$ are, 
in principle, calculable in a given model: $f_{B}$ by solving the 
non-linear scalar field equations governing the formation of B-balls from the original 
Affleck-Dine condensate and $T_{d}$ by calculating the charge and decay rate of the B-balls accurately. $T_{d}$, which will depend explicitly on the reheating temperature 
after inflation, is the more model-dependent of the two.
 The reheating temperature can be estimated under the assumption 
that the baryon asymmetry originates from an Affleck-Dine 
condensate with CP violating phase of the order of 1, and, indeed, 
can be calculated given all the details of an inflation model, 
but $T_{R}$ is likely remain an important source of theoretical 
uncertainty in the B-ball decay scenario. However, it is quite possible that 
$T_{d}$ and $T_{R}$, by being sufficiently small and large relative 
to $T_{f}$ respectively, play no direct role in determining the final 
LSP density. 

The B-ball decay scenario for MSSM dark matter is a natural alternative to 
the thermal relic LSP scenario, and has the considerable advantage 
of being able to explain the similarity of the baryon and dark matter 
densities. Should future experimental constraints on the parameters of 
the MSSM prove to be incompatible with thermal 
relic dark matter but consistent with B-ball decay dark matter for 
some set of B-ball parameters, it would strongly support the 
B-ball decay scenario. In particular, should the LSP mass be 
determined experimentally, the ratio of the number density 
baryons to dark matter would then be constrained by nucleosynthesis 
bounds on the baryon asymmetry. This would impose significant 
constraints on the reheating temperature and B-ball parameters, which, 
by comparing with the theoretical value of $f_{B}$, could even 
provide a "smoking gun" for the validity of the B-ball decay 
scenario, should annihilations happen to play no role in 
determining the present LSP density.

\section*{Acknowledgments}
I should like to thank John McDonald for many useful discussions on Q-balls
and for an enjoyable collaboration. This work has been supported by the
Academy of Finland under the contract 101-35224.

\end{document}